\documentclass[11pt,english]{article}

\usepackage[T1]{fontenc}
\usepackage[latin9]{luainputenc}
\usepackage[a4paper]{geometry}
\usepackage{mathrsfs}
\usepackage{amsbsy}
\usepackage{amstext}
\usepackage{amssymb}
\usepackage{setspace}
\usepackage{esint}
\makeatletter
\newcommand{\lyxaddress}[1]{
\par {\raggedright #1
\vspace{1.4em}
\noindent\par}
}

\renewcommand\refname{New References Header} 

\makeatother

\usepackage{babel}
\begin{document}

\title{\textsuperscript{}Non-Cauchy surface foliations and their expected
connection to quantum gravity theories}

\author{Merav Hadad \\
{\small{}\hspace{0.2in} }}
\maketitle

\lyxaddress{{\small{}Department of Natural Sciences, The Open University of Israel
P.O.B. 808, Raanana 43107, Israel.
\\ E-mail: meravha@openu.ac.il }}
\begin{abstract}
We argue that quantum gravity theories should involve constructing
a quantum theory on non-Cauchy hypersurfaces and suggest that the
hypersurface direction should be the same as the direction of the
effective non-gravitational force field at a point. We start with a
short review of works which support this idea and then we foliate
spacetime along an effective non-gravitational force field direction.
Next we discuss the implication of this foliation on the expected
properties of quantum gravity. We also discuss the vagueness caused
by constructing any quantum theory when using non-Cauchy foliation. 
\end{abstract}

\section{Introduction}

Obtaining a gravitational theory from microscopic objects or quantum
fields is extremely important and challenging. It is important because
it is expected to help us understand problems of gravity at very high
energy, such as the behavior of black holes and the origin of the
universe. It is challenging because this kind of theory is as yet
unknown. Attempts to describe gravity using microscopic objects, such
as strings, loops or triangles, have not yet led us to the desired
Einstein equations. Whereas attempts to treat the gravitational metric
as simply another quantum field are problematic. They result in a
theory that involves spin-2 massless fields and this kind of theory
is not renormalizable in more than $2+1$ dimensions. Moreover, combining
quantum mechanics with general relativity is also problematic since
time plays a different role within these two frameworks: whereas general
relativity treats time as a dynamical variable, quantum theories use
the Hamiltonian formalism which causes time to act as an independent
parameter through which states evolve. Moreover, attempting to use
the Hamiltonian formalism in order to quantize the gravitational theory,
leads to a non renormalizable theory and to the problem of time in
the ADM formalism. 

Instead of giving up the powerful Hamiltonian formalism when general
relativity theories are concerned, we suggest to use this formalism
differently. We suggest to consider the symmetry breaking caused by
an effective non-gravitational force field, and to use its direction
as an independent parameter through which states evolve. This mean
that instead of singling out the direction of a time vector field
in the Hamiltonian formalism, we single out the direction of the effective
non-gravitational force. 

As we will show in the next section this idea is supported by several
works which all single out the direction of a non-gravitational force
in order to obtain different aspects of quantum gravity. It is also
in agreement with holography and Verlinde's suggestion. Moreover,
not only it uses the natural symmetry breaking caused by an effective
non gravitational force field, it also eliminates the problem of time
in the ADM formalism. Finally, this foliation enables us to obtain
the conditions which are needed in order to derive an effective (2+1)D
gravitational theory instead of (3+1)D theory, with can be renormelized
in the quantum limit. 

However, since the direction of any force field is space-like, and
not time-like this suggestion leads to vagueness regarding the basic
concepts of relativistic quantum field theories. We suggest ways to
overcome some of these vagueness in the last section.

This paper is organized as follows: we start with a review of works
which support the idea that quantum gravity theories should involve
constructing a quantum theory on non-Cauchy hypersurfaces and that
the hypersurface direction at any point should be the same as the
direction of the effective non-gravitational force field at this point.
Then we foliate spacetime along this direction, and discus the implication
of this foliation on the expected properties of quantum gravity. Finally,
we discuss the vagueness caused by constructing any quantum theory
when using non-Cauchy foliation, and discuss the implications on quantum
gravity. 

\section{Review on the relevance of Non-Cauchy Foliation to Quantum Gravity}

The main idea of this paper is that in order to find the proper way
to quantize the gravitational theory, we need to consider the symmetry
breaking caused by an effective non-gravitational force field, and
to use the direction of that field as the direction through which
states evolve. This suggestion is supported by several examples which
relate non-Cauchy surfaces to different aspects of quantum gravity. 

Before listing these examples, we first mention three known properties
relating quantum gravity to light-like hypersurfaces. The first is
the holographic principle \cite{holografic} , first proposed by 't
Hooft \cite{t Hooft}, which states that in quantum gravity, the description
of a volume of space can be encoded on a lower-dimensional boundary
to the region. Next is the AdS/CFT \cite{ADS/CFT} correspondence,
which uses a non-perturbative formulation of string theory to obtain
a realization of the holographic principle. And finally, the third
is Verlinde's suggestion \cite{Verlinde} that gravity is an entropic
force which arises from the statistical behavior of microscopic degrees
of freedom encoded on a holographic screen. We discuss the way these
are related to our suggestion and reinforce it, in the next section. 

However, in these descriptions the holographic screen is a light-like
surface and therefore a Cauchy surface. Conversely, treating any force
direction as the direction which states ''evolve'', involves non-Cauchy
hypersurfaces. Fortunately, non-Cauchy hypersurfaces in general, and
that resulting from singling out the direction of a non-gravitational
force in particular, have also been found useful when dealing with
aspects of quantum gravity. We will discuss in detail some such foliation.

\subsection{The thermodynamics properties of gravity can also be derived by non-Cauchy
foliation}

The first example we discus in details involves the surface density
of space time degrees of freedom (DoF). These are expected to be observed
by an accelerating observer in curved spacetime, i.e. whenever an
external non-gravitational force field is introduced \cite{Padmanabhan}.
This DoF surface density was first derived by Padmanabhan for a static
spacetime using thermodynamic considerations. We found that this density
can also be constructed from specific canonical conjugate pairs as
long as they are derived in a unique way. Here we briefly summaries
the assumptions and conclusions, for the detailed calculation see
\cite{merav}. 

To see this one starts by defining the direction of the space-like
vector field in a stationary D-dimensional spacetime. We take accelerating
detectors that have a $D$ velocity unit vector field $u^{a}$ and
acceleration $a^{a}=u^{b}\nabla_{b}u^{a}\equiv an^{a}$ (where $n^{a}$
is a unit vector and $u^{a}n_{a}=0$)\footnote{We assume that both unit vectors: $n^{a}$and $u^{a}$ are hyper surface
orthogonal and thus fulfill Frobenius's theorem}. Next one foliate spacetime with respect to the unit vector field
$n_{a}$ by defining a $(D-1)$- hyper-surface which is normal to
$n_{a}$. The lapse function $M$ and shift vector $W_{a}$ satisfy
$r_{a}=Mn_{a}+W_{a}$ where $r^{a}\nabla_{a}r=1$ and $r$ is constant
on $\Sigma_{D-1}$. The $\Sigma_{D-1}$ hyper-surfaces metric $h_{ab}$
is given by $g_{ab}=h_{ab}+n_{a}n_{b}$. The extrinsic curvature of
the hyper-surfaces is given by $K_{ab}=-\frac{1}{2}\mathcal{L}_{n}h_{ab}$
where $\mathcal{L}_{n}$ is the Lie derivative along $n^{a}$. The
$\Sigma_{D-2}$ hyper-surfaces metric $\sigma_{ab}$ is given by $h_{ab}=\sigma_{ab}-u_{a}u_{b}$. 

As was first noted by Brown \cite{Brown:1995su} for generalized theories
of gravity, the canonical conjugate variable of the extrinsic curvature
$K_{bc}$ is $4\sqrt{-h}n_{a}n_{d}U_{0}^{abcd}$. Where $U_{0}^{abcd}$
is an auxiliary variable, which equals $\frac{\partial\mathcal{L}}{\partial R_{abcd}}$
when the equations of motion hold. We found in \cite{merav} that
the relevant phase space for detectors with D-velocity $u^{a}$ at
point $P$ can be identified by projection of the extrinsic curvature
tensor and its canonical conjugate variable on the vector field $u^{a}$:\footnote{This means that we distinguish these canonically conjugate variables
from the others by projecting the extrinsic curvature and its canonical
conjugate variable along the time like unit vector $u_{b}$. Actually
this should be done more carefully since the Lie derivative of the
normal vector $u_{b}$ does not vanish in general and thus leads,
for example, to a contribution to the canonical conjugation of $h_{ab}$.}

\begin{eqnarray}
\left\{ K^{nm}u_{m}u_{n},4\sqrt{h}U_{0}^{abcd}n_{a}u_{b}u_{c}n_{d}\right\} %.\label{canonicalstring}
\label{tensor canonical tern}
\end{eqnarray}
The gravitational density degrees of freedom detected by an accelerating
detector with D-velocity $u^{a}$ at point $P$ is constructed from
multiplying these special stationary canonically conjugate variables.
Thus, using $K^{ab}u_{b}u_{a}=n^{a}a_{a}=a$, the gravitational $D-2$
surface density of the spacetime DoF observed by an accelerating observer
$\Delta n$ per unit time $\Delta t$ is
\begin{eqnarray}
\frac{\Delta n}{\Delta t} & = & 4a\sqrt{h}U_{0}^{abcd}n_{a}u_{b}u_{c}n_{d}.\label{density per time}
\end{eqnarray}

In order to derive $\Delta n$ one integrate this term during some
natural period of time. In this case we can use the assumption that
an accelerated observer detects her environment as an equilibrium
system at temperature $T$ even in curved spacetime\cite{Padmanabhan:2010xh}.
This kind of canonical ensemble gives thermal Greens functions which
are periodic in the Euclidean time with period $\beta=1/T$ \cite{B=000026D}.
Assuming this is also the case for fields which represent the spacetime
degrees of freedom we integrate eq. (\ref{density per time}) over
the Euclidian period $\beta$ and find that the surface density of the spacetime degrees
of freedom: 
\begin{eqnarray}
\Delta n=4\beta\sqrt{h}U_{0}^{abcd}a_{a}u_{b}u_{c}n_{d}.\label{canonical string}
\end{eqnarray}
Next, using the assumption that the temperature $T=1/\beta$ seen
by an accelerated observer equals $Na/2\pi$ even in curved spacetime
\cite{Padmanabhan:2010xh}, and $\sqrt{h}=N\sqrt{\sigma}$ ,we deduce
that the degrees of freedom surface density in a one period Euclidean
time detected by an accelerating observer is: 

\begin{eqnarray}
 &  & \Delta n=8\pi U_{0}^{abcd}\epsilon_{a}{}_{b}\mathbf{\mathbf{\mathbf{\boldsymbol{\epsilon}}}_{c}{}_{d}}.\label{canonical string1-1-1-1-2-2}
\end{eqnarray}
(Where $\epsilon_{ab}=\frac{1}{2}(n_{a}u_{b}-n_{b}u_{a})$ and $\mathbf{\boldsymbol{\mathbf{\epsilon}}_{ab}}=\sqrt{\sigma}\epsilon_{ab}$).
Up to factor of 4, this is precisely the expression for entropy density
of spacetime found by Padmanabhan using the equipartition law for
a static metric. 

To conclude: we see that entropy density of spacetime can also be
constructed from specific canonical conjugate pairs as long as they
are derived by foliating spacetime with respect to the direction of
the non-gravitational vector force field. That this aspect reinforces
the importance of singling out a very unique spatial direction: the
direction of a non-gravitational force.

\subsection{The affects of string vibrations on the metric can also be derived
by non-Cauchy foliation}

The second example we discuss in details involves string theory excitations.
String theory succeeds in describing a kind of quantum gravity, and
with its aid one can identify the entropy of a BPS black hole as string
theory excitations \cite{Strominger:1996sh,Mathur:2005zp}. Some of
these vibrations create metrics with quantized conical singularities
in the $r-y$ surface \cite{Giusto:2012yz}. Thus we expect that a
BPS black hole metric should be a kind of superposition of metrics,
some of which have a conical singularity in the $r-y$ surface. This
creates an uncertainty at the opening angle at the $r-y$ surface.
It turns out that these singularities can also be explained using
the uncertainty principle, as long as the variables in the uncertainty
principle are obtained in a unique way. Here we briefly summaries
the assumptions and conclusions for the detailed calculation see \cite{meravLevy}.

In order to see this we need to begin with the same foliation as before.
Only now the $\Sigma_{D-1}$ hypersurfaces metric $h_{ab}$ is given
by $g_{ab}=h_{ab}+(-1)^{s_{u}}u_{a}u_{b}$, where $s_{u}=0$ if $u_{a}$
is space-like and $s_{u}=1$ if $u_{a}$ is time-like. We also define
a $D-2$ hypersurfaces defined by $t=const$ and $r=const$ are thus
normal to the given vector $n_{a}$ and to $u_{a}$. The D-2 hypersurfaces
metric $\sigma_{ab}$ is given by $h_{ab}=\sigma_{ab}+(-1)^{s_{n}}n_{a}n_{b}$,
where $s_{n}=0$ if $n_{a}$ is space-like and $s_{n}=1$ if $n_{a}$
is time-like\footnote{In addition $n_{a}$ and $u_{a}$ must fulfill Frobenius's theorem.
All the $n_{a}$ and $u_{a}$ used in the examples on this paper fulfill
this theorem.}. Moreover, as was found out in \cite{meravLevy} , for the $D1D5$
black hole, instead of one unit normal vector $n_{a}$, one should
deal with a several vectors. It is a special case where we can find
$n$ ($n<D-1$) unit vectors: $n_{(i)a}$ ($i=1...n$) which are normal
to each other and to the same unit vector $u_{a}$ which is defined
by $n_{(i)}^{b}\nabla_{b}n_{(i)a}=a_{(i)}u_{a}$ and $a_{(i)}\neq0$.
In this case the $D-n-1$ metric $\tilde{\sigma}_{ab}$ is defined
by $h_{ab}=\tilde{\sigma}_{ab}+\sum_{i=1}^{n}(-1)^{si}n_{(i)a}n_{(i)b}$
and one obtains (see details in \cite{meravLevy}):
\begin{eqnarray}
\left[K^{ab}n_{(i)b}n_{(i)a}(x),\sqrt{{\sigma_{(i)}}}U_{0}^{abcd}u_{a}n_{(i)b}n_{(i)c}u_{d}(\tilde{x})\right]=\hbar\delta^{D-1}(x-\tilde{x})\label{canonical string-1}
\end{eqnarray}
where $\sqrt{{\sigma_{(i)}}}=\prod_{j\neq i}N_{j}\sqrt{-\tilde{\sigma}(-1)^{s_{u}+\sum_{l}sl}}$
.

For a $D1D5$ black hole we examine the static metric 
\begin{eqnarray}
ds^{2}=f(r)(-dt^{2}+dy^{2})+f(r)^{-1}(dr^{2}+r^{2}d\Omega^{2})+g(r)dz_{i}^{2},\label{Static spherical}
\end{eqnarray}
where $y$ and $z_{i}$ are compact. We will use the 'naive' geometry
\cite{Mathur:2005zp} where: $f(r)=(1+\frac{Q_{1}}{r^{2}})^{-1/2}(1+\frac{Q_{5}}{r^{2}})^{-1/2}$
and $g(r)=(1+\frac{Q_{1}}{r^{2}})^{1/2}(1+\frac{Q_{5}}{r^{2}})^{-1/2}$.

Choosing the two vectors: 
\begin{eqnarray}
n_{(1)}^{a}=(f(r)^{-1/2},0,0,....,0)\label{Static spherical}\\
n_{(2)}^{a}=(0,f(r)^{-1/2},0,0,....,0)\nonumber 
\end{eqnarray}
and define a $D-3$ matric $\tilde{\sigma}_{ab}$ as $h_{ab}=\tilde{\sigma}_{ab}+(-1)^{s1}n_{(1)a}n_{(1)b}+(-1)^{s(2)}n_{(2)a}n_{(2)b}$
, then
\begin{eqnarray}
\left[K_{tt}(x),N\sqrt{\tilde{\sigma}}U_{0}^{abcd}u_{a}n_{(1)b}n_{(1)c}u_{d}(\tilde{x})\right]=\hbar\delta^{D-1}(x-\tilde{x}).\label{action part1}\\
\left[K_{yy}(x),N\sqrt{\tilde{\sigma}}U_{0}^{abcd}u_{a}n_{(2)b}n_{(2)c}u_{d}(\tilde{x})\right]=\hbar\delta^{D-1}(x-\tilde{x}).
\end{eqnarray}
where $K_{tt}\equiv K^{ab}n_{(1)b}n_{(1)a}$, $K_{yy}\equiv K^{ab}n_{(2)b}n_{(2)a}$
and $N=N_{t}=N_{y}$. Integrating over a closed $D-1$ hypersurface
we find 
\begin{eqnarray}
\left[\Theta_{r-t_{E}},S_{Wr-t}\right]=\hbar\label{action part2}\\
\ \left[\Theta_{r-y},S_{Wr-y}\right]=\hbar
\end{eqnarray}
where\\
 $\Theta_{r-t_{E}}\equiv\oint ds_{1}K_{tt}=\oint Ndt_{E}K_{tt}$ is
the opening angle at the $r-t_{E}$ surface,\\
$S_{Wr-t}\equiv\oint Ndyd^{D-3}x\sqrt{\tilde{\sigma}}u_{a}n_{(1)b}n_{(1)c}u_{d}U_{0}^{abcd}$
\\
 $\Theta_{r-y}\equiv\oint ds_{2}K_{yy}=\oint NdyK_{yy}$ is the opening
angle at the $r-y$ surface,\\
 $S_{Wr-y}\equiv\oint Ndt_{E}d^{D-3}x\sqrt{\tilde{\sigma}}u_{a}n_{(2)b}n_{(2)c}u_{d}U_{0}^{abcd}$.

Finally we get the following uncertainty relation on the horizon of
a $D1D5$ black hole: 
\begin{eqnarray}
\Delta\Theta_{r-t_{E}}\Delta S_{W}\geq\hbar\label{action part2}\\
\Delta\Theta_{r-y}\Delta S_{Wr-y}\geq\hbar
\end{eqnarray}
The uncertainty at the opening angle in the $r-y$ surface for $D1D5$
black hole is in agreement with the fuzzball proposal, where it was
found that some specific string vibrations form metrics with conical
singularities at the $r-y$ surface \cite{Mathur:2005zp,Giusto:2012yz,Lunin:2002iz,Balasubramanian:2000rt}. 

To conclude: we see that these singularities can be explained not
only by string vibrations but also from quantum properties: the uncertainty
principle. Note that the variables in the commutation relations must
be canonical conjugate pairs which are obtained by singling out the
radial direction. The radial direction can be regarded as the direction
of a non gravitational force that causes observers to ``stand steel''
in these coordinate frame. Thus, these singularities, which according
to string theory are expected in quantum gravity theories, are derived
by the uncertainty principle only when singling out the non-gravitational
force direction. 

\subsection{Several expected quantum black hole properties can also be derived
by non-Cauchy foliation}

Finally lets mention shortly several example regarding the benefit
of the non-Cauchy surface foliation to the research of quantum black
hole. 

The first example comes long ago from the membrane paradigm \cite{Membrane}.
The membrane models a black hole as a thin, classically radiating
membrane vanishingly close to the black hole's event horizon. It turns
out that this non-Cauchy surface is useful for visualizing and calculating
the effects predicted by quantum mechanics for the exterior physics
of black holes. 

The second example involves the Wheeler-De Witt metric probability
wave equation. Recently, in \cite{Ghaffarnejad}, foliation in the
radial direction was used to obtain Wheeler-De Witt metric probability
wave equation on the apparent horizon hypersurface of the Schwarzschild-de
Sitter black hole. By solving this equation, the authors found that
a quantized Schwarzschild-de Sitter black hole has a nonzero value
for the mass in its ground state. This property of quantum black holes
leads to stable black hole remnants, which is also an expected property
of quantum gravity.

The last example we mention shortly for the benefit of non-Cauchy
foliation to quantum gravity involves \textquotedblleft holographic
quantization\textquotedblright . The holographic quantization uses
spatial foliation in order to quantize the gravitational fields for
different backgrounds in Einstein theory. This is carried out by singling
out one of the spatial directions in a flat background \cite{Park1},
and also singling out the radial direction for a Schwarzschild metric
\cite{park2}. Moreover, other works \cite{park3} even suggest that
the holographic quantization causes the (3+1)D Einstein gravity to
become effectively reduced to (2+1)D after solving the Lagrangian
analogues of the Hamiltonian and momentum constraints.

.

All these examples support the idea that quantum gravity theories
should involve constructing a quantum theory on non-Cauchy hypersurfaces
and that the hypersurface direction should be the same as the direction
of the effective non-gravitational force field at a point. In the
next section we foliate spacetime along this direction, and discuss
the implication of this foliation on Einstein equations the expected
properties of quantum gravity. 

\section{Expected Advantages of Non-Cauchy Foliation to Quantum Gravity.}

In order to discuss the implication of the foliation of spacetime
along the direction of an effective non-gravitational force field
to the expected properties of quantum gravity we need first to describe
properly some terms. In this section we foliate spacetime along this
direction and then discuss its expected benefits regarding the problems
of quantum gravity.

\subsection{Foliate spacetime along the direction of an effective non-gravitational
force field}

In order to single out the direction of the gravitational force, we
first define its direction. We define it as follow: for a given metric
in a D-dimensional spacetime, we take accelerating detectors that
have a $D$ velocity unit vector field $u^{a}$ and acceleration $a^{a}=u^{b}\nabla_{b}u^{a}\equiv an^{a}$
(where $n^{a}$ is a unit vector and $u^{a}n_{a}=0$). Note also that
this direction is a space-like vector field. Next, we foliate spacetime
with respect to the unit vector field $n_{a}$ by defining a $(D-1)$
hyper-surface which is normal to $n_{a}$ and constant $r$ on $\Sigma_{D-1}$.
The lapse function $M$ and shift vector $W_{a}$ satisfy $r_{a}=Mn_{a}+W_{a}$
where $r^{a}\nabla_{a}r=1$ and $r$ is constant on $\Sigma_{\text{r}}$.
The $\Sigma_{r}$ hyper-surfaces metric $h_{ab}$ is given by $g_{ab}=h_{ab}+n_{a}n_{b}$.
The extrinsic curvature of the hyper-surfaces is given by $K_{ab}=-\frac{1}{2}\mathcal{L}_{n}h_{ab}$
where $\mathcal{L}_{n}$ is the Lie derivative along $n^{a}$. The
intrinsic curvature $R_{ab}^{(3)}$ is then given by the 2+1 Christoffel
symbols: $\Gamma_{ab}^{k}=\frac{1}{2}h^{kl}\left(\frac{\partial h_{lb}}{\partial x^{a}}+\frac{\partial h_{al}}{\partial h^{b}}-\frac{\partial h_{ab}}{\partial x^{l}}\right)$
so that $R_{ab}^{(3)}=\frac{\partial\Gamma_{ab}^{k}}{\partial x^{k}}-\frac{\partial\Gamma_{ak}^{k}}{\partial x^{b}}+\Gamma_{ab}^{k}\Gamma_{kl}^{l}-\Gamma_{al}^{l}\Gamma_{lb}^{k}$
. Thus, instead of 3+1 Einstein equation,

\[
R_{ab}^{(4)}=8\pi\left(T_{ab}-\frac{1}{2}Tg_{ab}\right),
\]
one finds \cite{book 3+1} that the 3+1 Einstein equation, can be
written in terms of the $\Sigma_{r}$ hyper-surfaces metric $h_{ab}$,
it's extrinsic curvature $K_{ab}$ and the intrinsic curvature $R_{ab}^{(3)}$
:
\[
R_{ab}^{(3)}+KK_{ab}-2K_{ai}K_{b}^{i}-M^{-1}\left(\mathcal{L}_{r}K_{ab}+D_{a}D_{b}M\right)=8\pi\left(S_{ab}-\frac{1}{2}\left(S-P\right)h_{ab}\right),
\]
\[
R^{(3)}+K^{2}-K_{ab}K^{ab}=16\pi P,
\]
\[
D_{b}K_{a}^{b}-D_{a}K=8\pi F_{a}.
\]
where $D_{a}$ is the 2+1 covariant derivatives, $S_{ab}=h_{ac}h_{ad}T^{cd}$
, $P=n_{c}n_{d}T^{cd}$ and $F_{a}=h_{ac}n_{d}T^{cb}$. 

Note that whenever the extrinsic curvature does vanish, even only
on the hypersurface $r_{0}$, this foliation leads to an interesting
observation. Though a (2+1)D gravitational theory is believed to be
a toy model for quantum gravity, this foliation suggests that a (3+1)D
gravitational theory can be regard as an \textquotedblleft evolution\textquotedblright{}
of a (2+1)D gravitational theory along a non-gravitational force direction.
Thus, given the fact that a renormalized (2+1)D quantum gravity theory
can be obtained, this leads to a (3+1)D quantum gravity originated
from a (2+1)D renormalized theory. Whether or not this renormalized
construction leads to an effectively renormalized gravitational theory
on the (3+1)D is remain to be seen.

\subsection{Implication of this foliation on the expected properties of quantum
gravity}

Now, lets discuss the implication of this foliation on the expected
properties of quantum gravity. We expect this foliation to contribute
in three main topics. The first relates the equivalence principle
and the insights of this equivalence in the quantum limit. The second
is regarding the connection between quantum gravity and holography.
Finally the third validates Verlinde's suggestion which relates the
origin of a third spatial dimension to the existence and direction
of acceleration in a gravitational force. We expend on these issues.

\subparagraph{1) The equivalence principle in the quantum limit. }

We argue that the foliation along the non gravitational force, is
related to the equivalence principle and can contribute impotent insights
of this equivalence in the quantum limit. To see this, first note
that any effective non-gravitational force causes the observers to
accelerate in a spatial direction and thus breaks the symmetry between
the three spatial directions. As shown by the Unruh effect, this symmetry
breaking is extremely important when dealing with quantum theories
in flat space. Given the equivalence principle, we expect the same
symmetry breaking to occur for accelerating observers in curved spacetime.
Though it is extremely complicated (and maybe even impossible) to
derive Unruh radiation in curved space, this means that accelerating
detectors are expected to detect a kind of Unruh radiation \cite{Unruh (1976)}
of a gravitational field as well. Moreover, as was shown in \cite{merav},
singling out the effective non-gravitational field enables identification
of the relevant canonical term and calculation of the expected surface
density DoF obtained by acceleration observers. Thus singling out
the effective non-gravitational force is expected to be a powerful
tool in identifying and even eliminating the extra quantum gravitational
fields obtained by accelerating observers in curved space. 

\subparagraph{2) The procedure reinforces the connection between quantum gravity
and holography.}

We argue that this procedure reinforces the connection between quantum
gravity and holography. This can be seen in Hamiltonian formalism,
when singling out the effective non-gravitational force direction
instead of a direction of time. In general the Hamiltonian formalism
singles out the time direction and thus the time act as an independent
parameter through which states evolve. We suggest to consider the
symmetry breaking caused by an effective non-gravitational force field,
and to use the direction of that field as the direction through which
states evolve. This suggestion supports holography because using the
Hamiltonian formalism in this case, leads to equations which determine
the evolution of the gravitational fields along the effective non-gravitational
force direction, i.e. along the space-like vector field $r_{a}$.
Moreover, they are determined not only by the Hamiltonian equation,
but also by the \textquotedbl{}initial\textquotedbl{} condition on
the unique non-Cauchy hyper-surface $r=r_{0}$. Thus when singling
out the effective non-gravitational force direction instead of a direction
of time, we expect all the information needed to describe the gravitational
field in the $r\neq r_{0}$ is encoded on a non-Cauchy hyper-surface,
as suggested by holography. 

\subparagraph{3) This procedure validates Verlinde's suggestion.}

This procedure validates Verlinde's suggestion which relates the origin
of a third spatial dimension to the existence and direction of acceleration
in a gravitational force. To clarify this, note that when we write
the 3+1 Einstein equation, in terms of the $\Sigma_{r}$ hyper-surfaces
metric $h_{ab}$, its extrinsic curvature $K_{ab}=-\frac{1}{2}\mathcal{L}_{n}h_{ab}$
and the intrinsic curvature $R_{ab}^{(3)}$ ,we find that if on $r=r_{0}$:
$K_{ab}=0$,\footnote{and also using gauge fixing in order to eliminate the term $M^{-1}\left(\mathcal{L}_{r}K_{ab}+D_{a}D_{b}M\right)$ }
then all the observers who are restricted to be on $r=r_{0}$ are
not accelerating. This is because $a=n^{a}a_{a}=K^{ab}u_{b}u_{a}=0$.
Thus in this case gravity can be describe as a $(2+1)D$ theory. On
the other hand, if on some $r=r_{0}$: $K_{ab}\neq0$, then all the
observers on the surface do accelerate and the $(3+1)D$ Einstein
equation cannot be reduced to any kind of$(2+1)D$ Einstein equation.
Thus this foliation reinforces Verlinde's suggestion relating the
direction of the acceleration direction to an extra $3^{th}$ spatial
dimension. 

\subsection{The implication of this foliation on the obstacles of quantum gravity}

Next we discuss the implication of this foliation on the possibility
of solving the obstacles of quantum gravity which are the problem
of time in the ADM formalism and the renormalization problem. 

\subparagraph{Eliminate the problem of time in the ADM formalism.}

Our suggestion is strongly related to a problem with time in the ADM
formalism \cite{ADM}. This problem results from the conceptual difficulty
that arises when attempting to combine quantum mechanics with general
relativity. It arises from the contrasting role of time within these
two frameworks. In quantum theories time acts as an independent parameter
through which states evolve, with the Hamiltonian operator acting
as the generator of infinitesimal translations of quantum states through
time. In contrast, general relativity treats time as a dynamical variable
which interacts directly with matter. The problem of time occurs because
in general relativity the Hamiltonian is a constraint that must vanish
in vacuum. However, in any ordinary canonical theory, the Hamiltonian
generates time translations. Therefore, we arrive at the conclusion
that \textquotedbl{}nothing moves\textquotedbl{} (\textquotedbl{}there
is no time\textquotedbl{}) in general relativity. This so-called \textquoteleft frozen
formalism\textquoteright{} caused much confusion when it was first
discovered, since it seems to imply that nothing happens in a quantum
theory of gravity.  

The problem of time vanishes if, instead of the ordinary formulation
which splits between one dimension of time and three dimensions of
space, one splits between one dimension of space and the 2+1 dimensions
of spacetime. This is because in this case, thou the new Hamiltonian
vanish, the theory does still depend on the time coordinate even in
the vacuum.

\subparagraph{Solving the Non-Renormalization Problem}

Moreover, if we consider a non-Cauchy foliation in the vacuum, then
the vanishing of the new Hamiltonian turns into an advantage. This
accrue because it leads to the possibility of describing the vacuum
case as a renormalized (2+1)D quantum gravity theory. To see this,
note that if instead of time direction, we foliate along one of the
spatial directions, we arrive at the conclusion that, instead of \textquotedbl{}nothing
changes along the time direction\textquotedbl{}, it seems that \textquotedbl{}nothing
changes along this spatial direction\textquotedbl{}. In other words
\textquotedbl{}there is no third spatial dimension\textquotedbl{}
and the (3+1)D gravitational theory becomes effectively a (2+1)D theory.
This is excellent from the point of view of quantum gravity since
it is possible to quantize a (2+1)D gravitational theory. 

\section{Remarks on the expected causality problem due to the non-Cauchy foliation }

This paper deals with the possibility of employing the Hamiltonian
formalism in a unique way. Instead of singling out the direction of
a time vector field, we single out the direction of the effective
non-gravitational force. We expect this to be a powerful tool in order
to obtain a quantum gravity theory that does not diverge in some cases,
and to understand the origin of its divergence in others. However,
since this involves singling out a spatial direction instead of a
time direction, this also leads to vagueness regarding the basic concepts
of relativistic quantum field theories.

Moreover, this is expected to suffer from problems such as causality,
probability, conservation and unitarity and thus all these properties
should be reexamined in this case. In order to learn how to overcome
these difficulties we need to set aside the gravitational theory and
focus on investigating known quantum field theories in flat spacetime.
We need to investigate the possibility of obtaining relativistic quantum
theories by singling out a spatial direction. This is expected to
be useful since in this case we know what are the quantum properties
that should be obtained and the fact that these theories do not have
any of the expected problems.

In order to do that we repeat the known process of quantization, but
instead of singling out time and thus Cauchy surface which can be
defined as an equal time surface $x_{0}=0$, we use a non-Cauchy surface.
A non-Cauchy surface can be defined by constant spatial coordinate
surface $x_{1}=0$. In this case we expect that the quantization will
be as follows. For a given Lagrangian density of a relativistic field
theory $\mathscr{L}(\phi(x),\partial_{\mu}\phi(x))$, we define a
new canonically conjugate momentum to the field variable $\phi(x)$:
$\Pi_{1}(x)=\frac{\partial\mathscr{L}}{\partial{\phi'}(x)}$ where
$\phi'=\partial_{1}\phi$. The new Hamiltonian density will be $\mathscr{H}_{1}=\Pi_{1}(x)\phi'-\mathscr{L}$.
Next we need to verify that the dynamical equation derived by Euler-Lagrange
equation, can be written in the new Hamiltonian form as $\phi(x)'=\{\phi(x),H_{1}\}$
and $\Pi_{1}(x)'=\{\Pi_{1}(x),H_{1}\}$ (where $H_{1}=\int d^{3}\widetilde{x}\mathscr{H_{1}}\equiv\int dx^{0}dx^{2}dx^{3}\mathscr{H_{1}}$).
For this purpose we need to assume equal $x_{1}$ bracket relations
between the field variable $\phi(x)$ and the new conjugate momentum
$\Pi_{1}(x)$. However, if we use equal $x_{1}$ canonical Poisson
bracket relations:$\{\phi(x),\phi(y)\}_{x^{1}=y^{1}}=0=\{\Pi(x),\Pi(y)\}_{x^{1}=y^{1}}$
and $\{\phi(x),\Pi(y)\}_{x^{1}=y^{1}}=\delta^{3}(\widetilde{x}-\widetilde{y})=\delta(x^{0}-y^{0})\delta(x^{2}-y^{2})\delta(x^{3}-y^{3})$
we get a non causal relation. Thus equal spatial coordinate bracket
cannot be defined by the ordinary canonical Poisson brackets. We must
identify correctly the equal $x_{1}$ classical bracket relations
between the field variable $\phi(x)$ and the new conjugate momentum
$\Pi_{1}(x)$, in order to find out whether the dynamical equation
derived by Euler-Lagrange equation can be derived by using the Hamiltonian-like
form. However, it seems that in order to identify correctly the equal
$x_{1}$ classical bracket relations between the field variable $\phi(x)$
and the new conjugate momentum $\Pi_{1}(x)$ we need to have an extension
of the canonical Poisson brackets in such a way that it will be causally
defined even when foliating spacetime to non-Cauchy surfaces. Unfortunately,
we do not have this kind of extension, and it may be that this kind
of extension can not even be derived. For a scalar field, we suggested
\cite{new paper}. a way to identify the equal $x_{1}$ canonical
relation without using extended Poisson bracket and found that these
relations are 
\begin{eqnarray}
\{\phi(x),\phi(y)\}_{x^{1}=y^{1}} & = & -i\int\frac{d^{3}\widetilde{k}}{(2\pi)^{3}}\epsilon(\pm K^{0},P_{x}^{2})\frac{1}{P_{x}}e^{i\widetilde{k}\cdot(\widetilde{x}-\widetilde{y})}\label{commutation relation8-1}\\
\{\Pi_{1}(x),\Pi_{1}(y)\}_{x^{1}=y^{1}} & = & -i\int\frac{d^{3}\widetilde{k}}{(2\pi)^{3}}\epsilon(\pm K^{0},P_{x}^{2})P_{x}e^{i\widetilde{k}\cdot(\widetilde{x}-\widetilde{y})}\\
\{\phi(x),\Pi_{1}(y)\}_{x^{1}=y^{1}} & = & 0.
\end{eqnarray}
These relations are causal on the hypersurface $x^{1}=y^{1}$ and
it was found that the equation of motion derived by them and the new
Hamiltonian leads to the expected equations of motion. This means
that deriving a relativistic quantum gravity theory by singling out
a spatial direction instead of a time direction is possible, at least
for scalars in Minkowski space. However, in order to prove this generally
one needs to extend this work to other cases such as different kind
of fields, and hopefully in curved spacetime.

\section{Summary}

In this paper we suggest to use the Hamilton formalism in a unique
way. We suggest to consider the symmetry breaking caused by an effective
non-gravitational force field, and to use its direction as an independent
parameter through which states evolve. 

After a short introduction, we examined in the second section several
works which support this idea. All these works, which comes from different
areas in physics, single out the direction of a non-gravitational
force in order to obtain different aspects of quantum gravity. We
extended our discussion in two different areas: thermodynamics and
string theory. In the third section we found that in some cases the
(3+1)D gravitational theory can be regard as an \textquotedblleft evolution\textquotedblright{}
of a (2+1)D gravitational theory along a non-gravitational force direction.
Then, given the fact that a renormalized (2+1)D quantum gravity theory
can be obtained, this leads naturally to a (3+1)D quantum gravity
which is originated from a renormalized theory. In the third section
we also explained how this idea is in agreement with general expected
concepts relating quantum gravity such as holography and Verlinde's
suggestion. Moreover, we discussed the option that this unique foliation
is useful in order to eliminate the problem of time in the ADM formalism.
Finally, in the forth section we discussed the causality problem expected
from such non-Cauchy foliation. In this context we mention our latest
work which solved the expected causality problem expected from such
non-Cauchy foliation for scalars in Minkowski space.

To conclude, it seems that singling out the direction of an effective
non-gravitational force field, instead of the direction of time, may
leads to a (3+1)D quantum gravitational theory that can be regard
as an \textquotedblleft evolution\textquotedblright{} of a (2+1)D
renormalized quantum gravitational theory along a non-gravitational
force direction. Whether or not this renormalized construction leads
to an effectively renormalized gravitational theory on the (3+1)D
is remain to be seen. However, even if this way of construction does
not lead to a renormalized quantum gravitational field theory in the
(3+1)D, this construction can be considered as a proof that our inability
to renormalized the (3+1)D theory is related directly to the existence
of non-gravitational forces in our everyday life. 

\textbf{Acknowledgments:} This research was supported by The Open University of Israel's Research Fund (grant no. 509565 ) 

\renewcommand\refname{Bibliography}

\end{document}